\RequirePackage[displaymath]{lineno}

\documentclass[prc,apc,twocolumn,showkeys,showpacs,preprintnumbers,superscriptaddress]{revtex4-1}

\usepackage{graphicx}
\usepackage{subfigure}
\usepackage{dcolumn}
\usepackage{url}
\usepackage[hypertex,pagebackref=false,breaklinks=true]{hyperref}

\newcolumntype{d}[1]{D{.}{.}{#1}}
\newcolumntype{.}{D{.}{.}{-1}}

\begin{document}



\title{Hyperfine Field and Hyperfine Anomalies of Copper Impurities in Iron}


\author{V.V.~Golovko}
\altaffiliation{Present address: Department of Physics, Queen's
University, Stirling Hall, Kingston, ON, Canada, K7L3N6}%
\affiliation{Instituut voor Kern- en Stralingsfysica, Katholieke Universiteit Leuven, Celestijnenlaan 200D, B-3001 Leuven, Belgium}%


\author{F.Wauters}%
\affiliation{Instituut voor Kern- en Stralingsfysica, Katholieke Universiteit Leuven, Celestijnenlaan 200D, B-3001 Leuven, Belgium}%

\author{S.~Cottenier}
\affiliation{Center for Molecular Modeling, Ghent University, Technologiepark 903, B-9052 Zwijnaarde, Belgium}

\author{M. Breitenfeldt}
\affiliation{Instituut voor Kern- en Stralingsfysica, Katholieke Universiteit Leuven, Celestijnenlaan 200D, B-3001 Leuven, Belgium}%

\author{V. De Leebeeck}
\affiliation{Instituut voor Kern- en Stralingsfysica, Katholieke Universiteit Leuven, Celestijnenlaan 200D, B-3001 Leuven, Belgium}%

\author{S. Roccia}
\affiliation{Instituut voor Kern- en Stralingsfysica, Katholieke Universiteit Leuven, Celestijnenlaan 200D, B-3001 Leuven, Belgium}%

\author{G. Soti}
\affiliation{Instituut voor Kern- en Stralingsfysica, Katholieke Universiteit Leuven, Celestijnenlaan 200D, B-3001 Leuven, Belgium}%

\author{M. Tandecki}
\affiliation{Instituut voor Kern- en Stralingsfysica, Katholieke Universiteit Leuven, Celestijnenlaan 200D, B-3001 Leuven, Belgium}%

\author{E. Traykov}
\affiliation{Instituut voor Kern- en Stralingsfysica, Katholieke Universiteit Leuven, Celestijnenlaan 200D, B-3001 Leuven, Belgium}%

\author{S. Van Gorp}
\affiliation{Instituut voor Kern- en Stralingsfysica, Katholieke Universiteit Leuven, Celestijnenlaan 200D, B-3001 Leuven, Belgium}%

\author{D.~Z\'akouck\'y}%
\affiliation{Nuclear Physics Institute, ASCR, 250 68 \v{R}e\v{z}, Czech Republic}%
\author{N.~Severijns}%
\email{nathal.severijns@fys.kuleuven.be} \affiliation{Instituut voor Kern- en Stralingsfysica, Katholieke Universiteit Leuven, Celestijnenlaan 200D, B-3001 Leuven, Belgium}%

\date{\today}


\begin{abstract}
A new value for the hyperfine magnetic field of
copper impurities in iron is obtained by combining resonance frequencies from
$\beta$-NMR/ON experiments on $^{59}$Cu, $^{69}$Cu and $^{71}$Cu with magnetic moment values from collinear laser
spectroscopy measurements on these isotopes. The resulting value, i.e. $B_{\rm{hf}}(\rm{Cu}{\it{Fe}})$~=~$-$21.794(10)~T,
is in agreement with the value adopted until now but is an order of magnitude more precise.
It is consistent with predictions from {\emph{ab~initio}} calculations. Comparing the hyperfine field
values obtained for the individual isotopes, the hyperfine anomalies in Fe were determined to be
$^{59}\Delta^{69}$ = 0.15(9)~\% and $^{71}\Delta^{69}$ = 0.07(11)~\%.

\end{abstract}





\maketitle



\section{\label{sec:intr}Introduction}

Precise values of magnetic hyperfine fields \cite{Rao85} allow the
determination of nuclear magnetic moments by experimental methods such
as integral perturbed angular correlation (IPAC) \cite{Bodenstedt75} and
time-differential perturbed angular distribution (TDPAD) \cite{Goldring85},
or low-temperature nuclear orientation (LTNO)
and nuclear magnetic resonance on oriented nuclei (NMR/ON) \cite{Stone86}.
Precise results, in addition, allow a detailed comparison with theory.

The magnetic hyperfine fields of substitutional impurities in bcc Fe are
at present well understood for most of the elements in the periodic table
\cite{Akai84,Akai85a,Akai85b,Korhonen00,Cottenier00,Torumba06,Torumba08}, with sizeable
differences between theory and experiment remaining mainly for the heavier
5\textit{d} impurities \cite{Ebert90,Severijns09}, the
alkaline elements \cite{Wouters87,Ashworth90,Vanderpoorten92,Will98},
and the actinides \cite{Torumba08}. Still,
for a few elements no precise experimental results are available yet.
A special case is copper, for which the currently accepted value of the hyperfine field
in iron is $\it{B}_{\rm{hf}}({\rm{Cu}}{\it{Fe}})$~=~$-$21.8(1)~T~\cite{Khoi75}.
This value resulted from a spin-echo NMR (nuclear magnetic
resonance) measurement with the sample containing the stable isotopes $^{63}$Cu
and $^{65}$Cu cooled cooled to a temperature of 4.2~K.
The error of 0.1~T was not given in the original publication
but was provided later as a private communication by
one of the original authors \cite{Rikovska2000a}. The sign was obtained
from the field shift in NMR measurements \cite{Kontani1967} and was
confirmed by theoretical calculations (Ref.~\cite{Akai85} and Sec.~\ref{theo}).


In the past few years the magnetic hyperfine interaction frequencies
$\nu_{\rm{res}} \sim \mu B_{\rm{tot}}$, with $\mu$ the nuclear magnetic moment
and $B_{\rm{tot}}$ the total magnetic field the nuclei experience, have been
determined for the Cu isotopes $^{59}$Cu \cite{Golovko04}, $^{67}$Cu
\cite{Rikovska2000a}, $^{69}$Cu \cite{Rikovska2000b}, and $^{71}$Cu
\cite{Stone08a} with the $\beta$-NMR/ON method,
i.e. NMR/ON with $\beta$-particle detection.
In these measurements, performed on samples that were cooled to millikelvin temperatures,
the above mentioned value for the hyperfine field of Cu impurities in
Fe host, viz. $B_{\rm{hf}}(\rm{Cu}{\it{Fe}})$~=~$-$21.8(1)~T,
was used in order to extract the nuclear magnetic moments for the
isotopes studied.
Recently, the magnetic moments of the Cu isotopes with A~=~61 to 75,
i.e. including the ones mentioned above, have been determined by collinear
laser spectroscopy measurements at ISOLDE-CERN
\cite{Flanagan09,Vingerhoets10,Vingerhoets11}.
As results for the $\beta$-NMR/ON resonance frequencies and for the
magnetic moments from the laser spectroscopy measurements
for these isotopes all have similarly high precisions, ranging from
2 to 6~$\times$~10$^{-4}$, combining these results allows determining
a new and precise value for the
hyperfine magnetic field of Cu in Fe at 0~K.

Note that the resonance frequencies and magnetic moment values that
will be used here were all obtained with the same experimental methods
and setups. This reduces the risk for possible systematic errors related to
calibration issues.

\section{Nuclear magnetic resonance on oriented nuclei}

The NMR/ON method has been applied widely for the
precise determination of the magnetic hyperfine splitting
of radioactive nuclei in the ferromagnetic host lattices
Fe, Co and Ni.
In most cases, the primary goal was to deduce the nuclear
magnetic moments of the impurity isotopes \cite{Stone05}.
The same technique is also used to ground state spins \cite{Vandeplassche86},
hyperfine fields \cite{Krane83,Rao85}, nuclear relaxation times
\cite{Bacon72,Herzog86}, and quadrupole splittings
\cite{Johnston72,Hagn85}, as well as to obtain information on the
lattice location and implantation behavior of implanted impurities
~\cite{Pattyn76,Dammrich88,vanWalle86,Severijns09}.
NMR/ON experiments require the nuclei to be oriented, which is
done using the LTNO method~\cite{Stone86} and requires cooling down
the radioactive samples to temperatures in the millikelvin region
and subjecting them to high magnetic fields, either hyperfine magnetic
fields \cite{Stone86} or externally applied fields \cite{Brewer86}.

In NMR/ON, the resonant depolarization of the radioactive probe
nuclei is detected as a function of the frequency of the applied
radio-frequency field via the resulting destruction of anisotropy
in the anisotropic emission of the decay radiation.
The resonance frequency is related to the hyperfine field, $B_{\rm{hf}}$,
through the relation
\begin{linenomath*}
\begin{equation}\label{eq:resFreqPractical}
    \nu_{\rm{res}}[\rm{MHz}]
    = \left| \frac { 7.6226 \cdot
      \mu[\mu_{\rm{N}}] \textit{B}_{\rm{tot}}[\rm{Tesla}] }
    {\textit{I}[\hbar]} \right|
\end{equation}
\end{linenomath*}
\noindent with $I$ the nuclear spin of the isotope studied and
\begin{equation}\label{eq:Btot}
B_{\rm{tot}}=B_{\rm{hf}}+B_{\rm{app}}(1+K)-B_{\rm{dem}} ~ .
\end{equation}
\noindent Here the hyperfine field $B_{\rm{hf}}$ includes the Lorentz field
of 0.742~T for bcc iron at 0~K, $B_{\rm{app}}$ is the externally applied
magnetic field, $B_{\rm{dem}}$ is the demagnetization field
and, $K$ is the Knight shift. The constant factor is the ratio of the
fundamental constants $\mu_N / h$ and does not contribute to the error budgets here.

\section{Collinear laser spectroscopy}

Collinear laser spectroscopy experiments determine the hyperfine structure
of atomic ground- and excited states, yielding precise values for the
hyperfine parameters $A$ and $B$, which in turn provide accurate values
for magnetic and quadrupole moments of the isotopes studied~\cite{Mueller83}.

Recently, collinear laser spectroscopy experiments were performed on a
series of Cu isotopes at ISOLDE with the COLLAPS setup, including the four isotopes for which resonance
frequencies are available from recent $\beta$-NMR/ON measurements.
For several of these isotopes laser spectroscopy could only be performed
after the installation of the ISCOOL
cooler and buncher radio-frequency quadrupole Paul trap \cite{Franberg08,Mane09}.
This allowed the collinear laser setup to be operated in bunched
mode, reducing the background photon counts from scattered laser
light by more than three orders of magnitude, and permitting to perform
measurements on isotopes that were previously not accessible due to their
low yields. The Cu$^+$ bunches from ISCOOL were sent through a sodium
vapor cell which neutralized the ions through charge-exchange collisions.
A voltage was applied to the vapor cell for tuning the velocity of the ions
and bringing them onto resonance with the laser beam that was overlapped
with the Cu beam in the co-propagating direction. Resonances were located
by measuring the photon yield as a function of the voltage with two
photomultiplier tubes, the voltage of which was gated so that photons
were only recorded when an atom bunch was within the light collection
region \cite{Flanagan09}.
An example of a collinear resonance fluorescence spectrum obtained in these
measurements is shown in Fig.~1 of Ref.~\cite{Flanagan09}.

\section{Hyperfine magnetic field for copper in iron}

In the following the magnetic moments from the collinear laser spectroscopy
measurements on the isotopes $^{59}$Cu \cite{Vingerhoets11}, $^{67}$Cu
\cite{Vingerhoets10}, $^{69}$Cu \cite{Vingerhoets10} and $^{71}$Cu \cite{Flanagan09}
are combined with the $\beta$-NMR/ON measurements that were performed
on these isotopes \cite{Golovko04,Rikovska2000a,Rikovska2000b,Stone08a}.
The experimental magnetic moment values and $\beta$-NMR/ON resonance frequencies
for $^{59}$Cu, $^{69}$Cu, and $^{71}$Cu are listed in Table~\ref{tab:Cu hf fields}
and will be discussed in Sec.~\ref{results-59-69-71}. The data for $^{67}$Cu
are discussed separately in Sec.~\ref{results-67}.

As the extraction of the hyperfine magnetic field using Eqs.~\ref{eq:resFreqPractical}
and \ref{eq:Btot} requires that a possible Knight shift and the demagnetization field
are taken into account as well, the $\beta$-NMR/ON measurements and these two factors
will be addressed in some detail first.

\subsection{$\beta$-NMR/ON measurements}

\begin{figure}[top]
 \centering
 \includegraphics[width=\columnwidth]{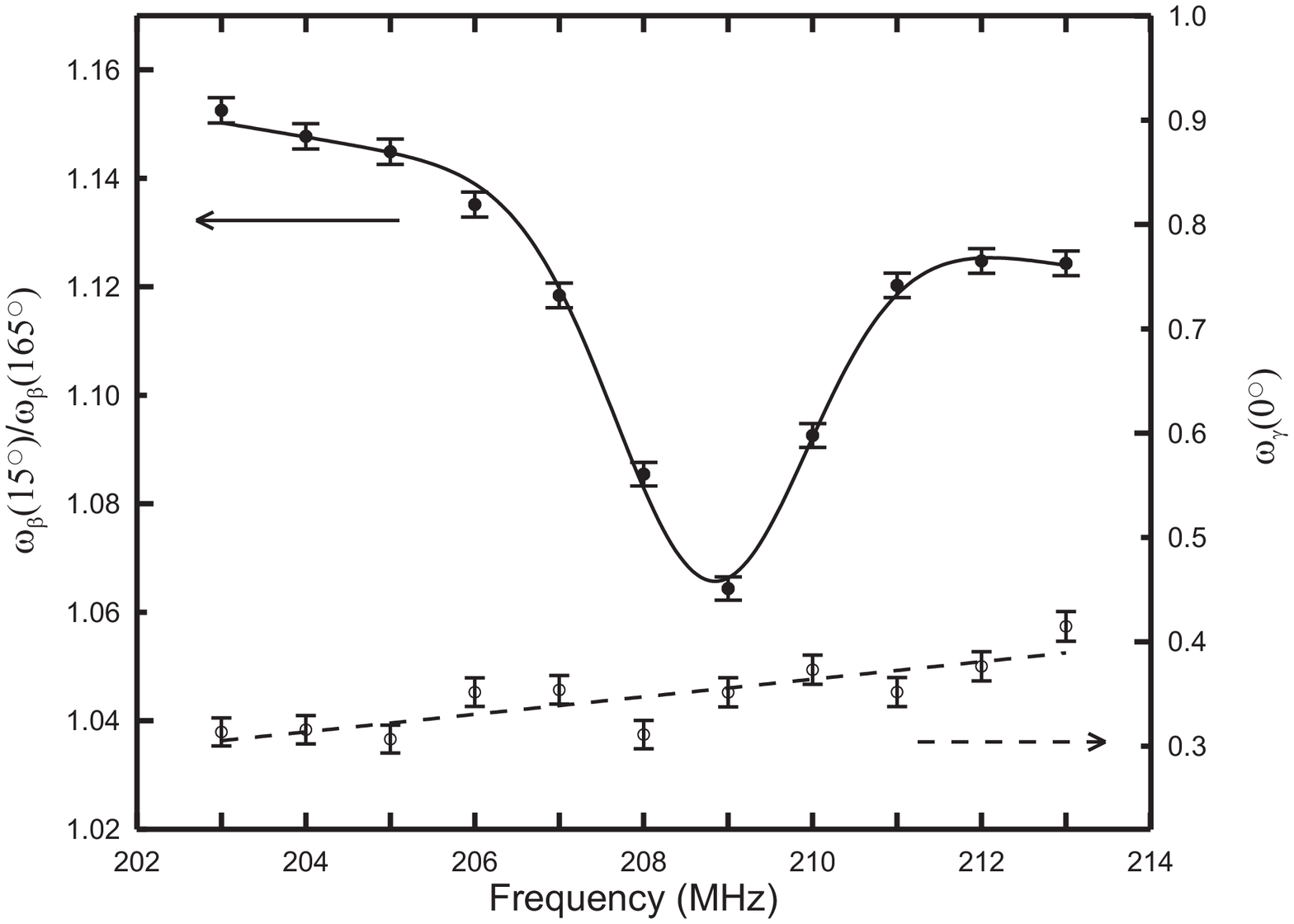}
 \caption{\label{fig:Beta_Res} On-line NMR/ON curve for $^{59}$Cu isotope.
 Plotted is the ratio of the count rates for two detectors at
 $15^{\circ}$ and $165^{\circ}$ with respect to the quantization
 axis as a function of frequency.
 At the bottom, the count rate at $0^{\circ}$
 is shown for the 136~keV $\gamma$~ray of
 the $^{57}\rm{Co}\it{Fe}$ thermometer
 (corresponding to a sample temperature of about 10 mK)
 for the same frequency region, showing no effect at the position of
 the $^{59}$Cu resonance. The slope in the count rates is
 caused by a small increase of the sample temperature with
 the resonance frequency. (arrows indicate the vertical scale that applies;
 from ref.~\cite{Golovko04})}
\end{figure}

All four Cu isotopes on which $\beta$-NMR/ON measurements have been performed
were produced at the ISOLDE isotope separator facility.
The NMR/ON measurements were performed either on-line using the NICOLE LTNO
setup \cite{Schlosser88} at ISOLDE (for $^{59}$Cu \cite{Golovko04}, $^{69}$Cu
\cite{Rikovska2000b} and $^{71}$Cu) \cite{Stone08a}, or off-line with the nuclear-orientation facility at Oxford University (for $^{67}$Cu \cite{Rikovska2000a}).
As an example, the NMR/ON result for $^{59}$Cu$\it{Fe}$ is shown in
Fig.~\ref{fig:Beta_Res}.

\subsubsection{Demagnetization field}
\label{demag}

The calculation of the demagnetization field is not straightforward,
but for the simple shapes and thin foils (magnetized in the plane of the foil)
that were used in the $\beta$-NMR/ON
measurements discussed here, analytical expressions
can be obtained \cite{Chikazumi64}.
At the current level of precision, the small demagnetization fields in
the foils used, which are of the order of 0.01~T, cannot be neglected.

The foil used by the Leuven group for the
measurement with $^{59}$Cu had a size
of 9~mm~by~14~mm and an initial thickness of 125~$\mu$m.
It was polished with diamond-base paste with grain sizes of 3~$\mu$m
and 1~$\mu$m. It is estimated that this reduced the thickness of this foil
to (95~$\pm$~20)~$\mu$m. For this a demagnetization field
$B_{\rm{dem}}$~=~0.020(4)~T is calculated.
The foils used by the Oxford group for the measurements with
$^{67}$Cu, $^{69}$Cu and $^{71}$Cu had an initial thickness of 250~$\mu$m
\cite{VanEsbroeck05}. They were first polished with sand paper with CAMI
(Coated Abrasive Manufacturers Institute) grit designation 600
(i.e. with an average particle diameter of 16~$\mu$m) and thereafter also
with diamond-base paste. Due to the larger particle diameter used
in the first step, this procedure removed
more material from the foils leading to an estimated thickness of
(190~$\pm$~40)~$\mu$m. For typical dimensions of 10~mm~by~15~mm
for such foils, a demagnetization field $B_{\rm{dem}}$~=~0.036(8)~T
is then calculated for the measurements with $^{67}$Cu, $^{69}$Cu and $^{71}$Cu.

\subsubsection{Knight shift}
\label{knight}

The Knight shift for copper in iron has never been determined at
low temperatures. For other elements in ferromagnetic host materials
Knight shift values ranging from zero to about 5\% have been reported
\cite{Yazidjoglou1993}.
Thus, for the low external fields, $B_{\rm{app}}$, of 0.1~T and 0.2~T that were used
in the $\beta$-NMR/ON measurements discussed here, the Knight shift corrections
could amount up to about 0.005~T and 0.010~T, respectively. These values
were then used as the 1$\sigma$ error bars on the values for the externally
applied magnetic fields. Note that the precision to which these external fields
could be set is an order of magnitude better and can therefore be
neglected at the current level of precision.

\subsection{Results for $^{59}$Cu, $^{69}$Cu and $^{71}$Cu}
\label{results-59-69-71}

The magnetic moment values and the resonance frequencies for the isotopes
$^{59}$Cu, $^{69}$Cu and $^{71}$Cu are listed in Table~\ref{tab:Cu hf fields}
(the case of $^{67}$Cu will be discussed separately in the next section).
Also listed there are the values for the total magnetic field, $B_{\rm{tot}}$,
obtained for each isotope using Eq.~\ref{eq:resFreqPractical}, as well as the
hyperfine field, $B_{\rm{hf}}$, that is then obtained from Eq.~\ref{eq:Btot}
using the values for the externally applied field, $B_{\rm{app}}$, and the
demagnetization field, $B_{\rm{dem}}$.

As can be seen, the hyperfine field values obtained for these three isotopes are
in very good agreement with each other (see also Fig.~\ref{fig:hyperfine fields}).
Combining all three results yields a weighted average value of
\begin{equation}
B_{\rm{hf}}({\rm Cu} \it{Fe})~=~\rm{-21.794(10)~T}  ~   .
\label{eq:average}
\end{equation}

The differences between the hyperfine field values obtained for the three isotopes
listed in Table~\ref{tab:Cu hf fields} could be due to small differences
in the distribution of the nuclear magnetization over the nuclear volume for the
three isotopes, i.e. hyperfine anomalies \cite{Butgenbach84}. The usual definition
for the hyperfine anomaly, $\epsilon_i$, for a single nuclear state $i$ is
\begin{equation}
B_{\rm{eff}} = B_{\rm{0}} (1 +\epsilon_i)  .
\end{equation}
\noindent where $B_{\rm{eff}}$ is the hyperfine field averaged over the distribution of
nuclear magnetization of the state $i$ and $B_{\rm{0}}$ is the hyperfine field at
$r$ = 0. The difference of the hyperfine anomalies of two nuclear states in the same host metal
is then given by
\begin{equation}
^{1}\Delta^{2} = \epsilon_1 - \epsilon_2 = \frac{B_{\rm{eff}}^1}{B_{\rm{eff}}^2} - 1   .
\end{equation}
\noindent
When considering the possible presence of hyperfine anomalies the values for
$B_{\rm{hf}}$ listed in Table~\ref{tab:Cu hf fields} have to be interpreted as values
for $B_{\rm{eff}}$. It then turns out that the differences of the hyperfine anomalies
of these three isotopes are less than 3~$\times$~10$^{-3}$
(see Table~\ref{tab:Cu hf fields}; 90\% C.L.). Note that
previously the hyperfine anomaly between $^{63}$Cu and $^{65}$Cu was estimated
to be less than 5~$\times$~10$^{-5}$ \cite{Locher74}, while in the recent collinear
laser spectroscopy measurements on Cu isotopes \cite{Vingerhoets10}, no indication
for hyperfine anomalies was observed either for the series of isotopes ranging
from $A$~=~61 to 75.

For $^{59}$Cu, two other values for the magnetic moment were recently quoted as well,
i.e. $\mu(^{59}$Cu)~=~1.83(4)~$\mu_N$, from in-source laser spectroscopy \cite{Stone08b},
and  $\mu(^{59}$Cu)~=~1.910(4)~$\mu_N$, from an in-gas-cell laser spectroscopy
experiment \cite{Cocolios09,Cocolios10}. As the first value is much less precise than the
one obtained from collinear laser spectroscopy it is not further used here.
The second value differs well outside error bars from the collinear laser spectroscopy result
(see Table~\ref{tab:Cu hf fields}).
Combining it with the $\beta$-NMR/ON resonance frequency of 208.79(4)~MHz for $^{59}$Cu
reported in Ref.~\cite{Golovko04},
a hyperfine field value of $B_{\rm{hf}}(\rm{Cu}\it{Fe}$)~=~$-$21.59(5)~T is obtained.
This differs by about 4 standard deviations from the values listed in
Table~\ref{tab:Cu hf fields} that resulted from combining
resonance frequencies and magnetic moment values that were
each obtained with the same experimental setups.
This difference might be the due to unforeseen systematic effects in this measurement
\cite{Cocolios09} which was performed with a different experimental setup,
as is currently being investigated \cite{Cocolios11}.

\subsection{The case of $^{67}$Cu}
\label{results-67}

\begin{figure}[top]
\centering
\includegraphics[width=\columnwidth]{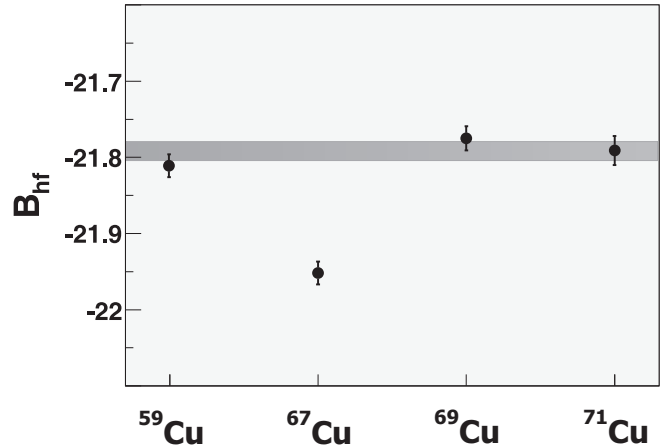}
\caption{\label{fig:hyperfine fields} Hyperfine field values (in Tesla) for Cu in Fe obtained by combining
resonance frequencies from $\beta$-NMR/ON measurements on the
isotopes $^{59}$Cu, $^{67}$Cu, $^{69}$Cu and $^{71}$Cu with the
magnetic moments for these isotopes from collinear laser spectroscopy. The grey band indicates the weighted average of the values from $^{59}$Cu, $^{69}$Cu and $^{71}$Cu.}
\end{figure}

The situation for $^{67}$Cu turns out to be more complicated.
In Ref.~\cite{Rikovska2000b}, the magnetic moment of $^{67}$Cu obtained with the Oxford LTNO setup
is given as $\mu(^{67}$Cu)~=~+2.54(2)~$\mu_N$; in Ref.~\cite{Rikovska1999}
this result is quoted as $\mu$~=~+2.536(3)~$\mu_N$ (this smaller error bar
most probably only reflects the statistical precision).
The authors do not quote a resonance frequency value, but analysis of the resonance curve shown
in Ref.~\cite{Rikovska2000b} yields a central value of $\nu_{\rm{res}}$~=~278.38(6)~MHz.
In combination with the magnetic moment value of $\mu$~=~+2.536~$\mu_N$, this
yields a total magnetic field of $-$21.60~T. Comparing this with the hyperfine field value of $-$21.8(1)~T
for copper in iron that was used in Ref.~\cite{Rikovska2000b} and neglecting the
demagnetization field (as the authors of Ref.~\cite{Rikovska2000b} also did),
yields a value of +0.20~T for the externally applied magnetic field instead
of the value of +0.10~T mentioned in Refs.~\cite{Rikovska1999,Rikovska2000b}.
The latter value is therefore most probably a typographical error.
\\
%
When combining the above resonance frequency of 278.38(6)~MHz
with the magnetic moment value $\mu(^{67}$Cu)~= +2.5142(6)~$\mu_N$ from the laser
spectroscopy experiments \cite{Vingerhoets10},
a hyperfine field value of $B_{\rm{hf}}(\rm{Cu}\it{Fe})$~= $-$21.952(15)~T
is obtained for $B_{\rm{ext}}$~=~0.20(1)~T and with $B_{\rm{dem}}$~= 0.036(8)~T
for the foils of the Oxford team (see Sec.~\ref{demag}). If the demagnetization
field is neglected $B_{\rm{hf}}(\rm{Cu}\it{Fe})$~= $-$21.988(15)~T
is obtained. Both values are at variance with the ones obtained for the other
Cu isotopes (see Table~\ref{tab:Cu hf fields} and Fig.~\ref{fig:hyperfine fields}).
\\
Concluding, there seems to be a problem with the $\beta$-NMR/ON
result for $^{67}$Cu that was obtained using another experimental setup than
the one used for the isotopes $^{59}$Cu, $^{69}$Cu and $^{71}$Cu.
The origin of this may e.g. be an undetected error in the frequency calibration.
We therefore did not include the
$\beta$-NMR/ON result for $^{67}$Cu in the hyperfine field analysis presented here.

%

%
%
\begin{table*}[htb]
\begin{center}
\caption{Data used for the determination of the hyperfine field of Cu impurities in Fe and values deduced from these. The error bar on the weighted average was increased by a factor $\sqrt{\chi^2/\nu}$~=~$\sqrt{1.22}$ to account for the fact that the reduced $\chi^2$ is larger than unity. The last column lists the hyperfine anomaly differences between $^{69}$Cu and the two other isotopes.
}
\label{tab:Cu hf fields}
%
\begin{tabular*}{\textwidth}{@{\extracolsep{\fill}}lccccccccc}

\hline \hline 

Isotope      &    $\mu$     &    Ref.                & $\nu_{\rm{res}}$ &   Ref.     & $B_{\rm{tot}}$\footnotemark[1] &   $B_{\rm{app}}$\footnotemark[2] & $B_{\rm{dem}}$ & $B_{\rm{hf}}$\footnotemark[1] & $^{X}\Delta^{69}$  \\
             &  ($\mu_N$)   &                        &       (MHz)      &                        &      (T)      &      (T)         &     (T)       &     (T)     &     (\%)  \\
    \hline
             &              &                        &                  &                        &               &                  &               &               \\

  $^{59}$Cu  & $+1.8910(9)$  & \cite{Vingerhoets11} &    $208.79(4)$   &  \cite{Golovko04}      & $-21.727(11)$  & $0.100(5)$       &  $0.020(4)$\footnotemark[3]  & $-21.807(13)$  & 0.15(9)  \\


  $^{69}$Cu  & $+2.8383(10)$ & \cite{Vingerhoets10}  &    $311.7(1)$    &  \cite{Rikovska2000b}  & $-21.611(10)$  & $0.20(1)$       &  $0.036(8)$\footnotemark[3]  & $-21.775(16)$  &   -  \\

  $^{71}$Cu  & $+2.2747(8)$ &   \cite{Flanagan09}   &   $250.00(14)$   &  \cite{Stone08a}       & $-21.627(14)$  & $0.20(1)$       &  $0.036(8)$\footnotemark[3]  & $-21.791(19)$   &  0.07(11)  \\

  \hline
  Weighted average    &         &                     &                  &                        &               &                  &     & $-21.794(10)$   \\

  \hline
  \hline

\end{tabular*}
\end{center}
\footnotemark[1]{Sign from \cite{Kontani67}.}
\footnotemark[2]{From the Refs. listed in column~3.}
\footnotemark[3]{Calculated with the formulas derived in Ref.~\cite{Chikazumi64} (see also Sect.~\ref{demag}).}
%
\end{table*}
\begin{table}[phtb]
\centering
\caption{Overview of hyperfine magnetic fields for Cu in Fe host reported in the literature.
Note that in most cases no error bar was quoted.}
\label{tab:other Cu hf fields}

\begin{ruledtabular}
\begin{tabular}{d{-1} l c}
 \multicolumn{1}{c}{$B_{\rm{hf}}$ (T)}  &      Ref. & method \\ [2pt]
    \hline
    \\ [-6pt]
21.0      &    \cite{Koi62}                    &   NMR          \\
21.77     &    \cite{Kushida62}                &   NMR          \\
-21.27    &    \cite{Shirley65}                &   NMR          \\
21.79     &    \cite{Edmonds66}                &   NMR          \\
-21.3     &    \cite{Kontani67}                &   spin-echo    \\
21.8(1)   &    \cite{Khoi75,Rikovska2000a}      &  NMR    \\
21.3      &    \cite{Riedi81}                  &   NMR          \\
-21.84    &    \cite{Kasamatsu86}              &   NMR          \\
16.95(51) &    \cite{Lohmann93}                &   $\gamma-\gamma$ PAC\footnotemark[1]  \\
\end{tabular}
\end{ruledtabular}
\footnotemark[1]{perturbed angular correlation.}
\end{table}

\section{Previous results}

Table \ref{tab:other Cu hf fields} lists other values for the hyperfine field of
Cu in Fe that are available in the literature, most of which were obtained in
classical NMR experiments at room temperature and with - unfortunately - no
error being quoted. As can be seen, most values are in reasonable agreement with
the new value of $-$21.794(10)~T presented here.
The value from the $\gamma$-$\gamma$ perturbed
angular correlation measurement, which is significantly deviating from all other
results, is either wrong or might be related to a different
lattice site for Cu impurities in Fe.

\section{Comparison with theoretical values}
\label{theo}

Hyperfine fields in solids can be calculated from first principles. This allows understanding trends in those hyperfine fields as a function of the impurity element in e.g. an iron matrix~\cite{Akai84,Akai85a,Akai85b,Korhonen00,Cottenier00,Torumba06,Torumba08},
and it allows disentangling the hyperfine field into contributions with a different physical origin~\cite{Novak03,Torumba06,Torumba08,Peltzer09}. We have calculated the hyperfine field of Cu in Fe within the framework of Density Functional Theory~\cite{Kohn64,Kohn65,DFT-LAPW02}, using the Perdew-Burke-Ernzerhof (PBE) exchange-correlation functional~\cite{Perdew96}. For solving the scalar-relativistic Kohn-Sham equations we employed the \mbox{Augmented Plane Waves }\mbox{+ local orbitals} (APW+lo) method~\cite{Sjostedt00,Madsen01,DFT-LAPW02} as implemented in the {\scshape wien2}k package~\cite{Wien99} for periodic solids. In this method the wave functions are expanded into spherical harmonics inside nonoverlapping atomic spheres of radius $R_{\mathrm{MT}}$ and in plane waves in the remaining space of the unit cell, i.e. the interstitial region. We took $R_{\mathrm{MT}}^{\mathrm{Fe}}=R_{\mathrm{MT}}^{\mathrm{Cu}}=$~2.30~a.u.~.
The plane wave expansion of the wave function in the interstitial region was truncated at a large value of $K_{\mathrm{max}}=8.0/R_{\mathrm{MT}}^{\mathrm{min}}=$~3.48~a.u.$^{-1}$,
which leads to very accurate values for the calculated hyperfine fields.
A  dense mesh of k-points, corresponding to a 20$\times$20$\times$20 mesh for a conventional bcc unit cell for Fe, was taken. Spin-orbit coupling was taken into account by a second variational step scheme~\cite{Koeling77} using a cutoff energy E$_{SO}$=~5.0~Ry. The substitutional Cu impurity was modeled by a 128-atom supercell, and all atoms in the supercell were allowed to adjust their positions due to the presence of the impurity. The lattice constant for the Fe-matrix was taken to be the equilibrium lattice constant for the PBE functional (2.8404~$\AA$). These settings allow for an excellent numerical convergence of the hyperfine field.

The results of the calculations are summarized in Table~\ref{tab:theo}.
\begin{table}[htb]
\centering
\caption{Calculated values for the Fermi contact field,
orbital hyperfine field, dipolar hyperfine field and atomic dipolar
hyperfine field for a Cu impurity in Fe. Spin and orbital
contributions to the atomic magnetic moment for Cu ($\mu_B$) and
distance between Cu and its first Fe-neighbor ($\AA$).}
\label{tab:theo}
%
%
%
%
%
%
\begin{ruledtabular}
\begin{tabular}{l d{-1} lc}
%
\multicolumn{2}{c}{B(T)}          &                   & \\ [2pt]
\hline
\multicolumn{1}{l}{$B_{\rm{F}}$}             & -25.32 & $\mu_{\rm{spin}}$ & 0.115 \\
\multicolumn{1}{l}{$B_{\rm{orb}}$}           &   0.60 & $\mu_{\rm{orb}}$ & 0.007\\
\multicolumn{1}{l}{$B_{\rm{dip}}$}           &  -0.01 & $d_{\rm{Cu-Fe}}$ & 2.478\\
\multicolumn{1}{l}{$B_{\rm{dip}}^{\rm{at}}$} &   0.00 & & \\ [2pt]
\end{tabular}
\end{ruledtabular}
\end{table}
As can be seen, the distance between the Cu impurity and its first eight Fe neighbours is expanded only slightly (0.75~\%) compared to the Fe-Fe distance of 2.460~$\AA$ in pure Fe. The dominant contribution to the hyperfine field is the Fermi contact term, caused by s-electron spin polarization, due to the small atomic (d-electron) magnetic moment at the Cu atom. This value can be further split into a core contribution due to 1s and 2s electrons ($-$8.54~T), a semi-core contribution by the 3s-electrons (+5.80~T) and a valence contribution by 4s electrons ($-$22.59~T).
The result for the Fermi contact hyperfine field can be compared with the $-$18.2~T that was obtained 25 years ago by the Korringa-Kohn-Rostoker Green's function method~\cite{Akai85a}. The orbital hyperfine field of Cu (0.60~T) is an order of magnitude smaller than the corresponding quantity in pure Fe, consistent with the very small atomic orbital magnetic moment of Cu.  All contributions to the hyperfine field, together with the Lorenz field (0.74~T), sum to a total hyperfine field of $-$23.99~T. Although this is about 2~T larger than the experimental value determined in this work, this deviation is state-of-the-art, and is due to inherent limitations of the chosen exchange-correlation functional.

\section{Conclusion}

Combining resonance frequencies for the isotopes $^{59}$Cu, $^{69}$Cu and $^{71}$Cu obtained
from $\beta$-NMR/ON measurements with the NICOLE LTNO setup at ISOLDE, with magnetic moment
values obtained for these isotopes in collinear laser spectroscopy measurements
at ISOLDE, the hyperfine field of Cu impurities in iron is found to be
$B_{\rm{hf}}(\rm{Cu}\it{Fe}$)~=~$-$21.794(10)~T. This value is in agreement with but
almost an order of magnitude more precise than the previously adopted value of $-$21.8(1)~T
and in good agreement with predictions from {\emph{ab initio}} calculations.
Interpreting the differences between the hyperfine field values obtained for the
individual isotopes to be due to hyperfine anomalies, the hyperfine anomalies in Fe for
the isotopes considered here were found to be smaller than 3~$\times$~10$^{-3}$ (90\% C.L.).

\section{Acknowledgement}

This work was supported by the Fund for Scientific Research
Flanders (FWO), project GOA/2004/03 of the K. U. Leuven, the Interuniversity Attraction Poles Programme, Belgian
State Belgian Science Policy (BriX network P6/23), and the grant LA08015
of the Ministry of Education of the Czech Republic.


\end{document}